\documentclass{pazh2col-utf}
\usepackage{graphicx}
\usepackage{amsmath}
\usepackage{amssymb}

\begin{document}

%\aap - Astron. Astrophys.
% \aapr - Astron. Astrophys. Rev.
%\aaps - Astron. Astrophys. Suppl. Ser.
% \aj - Astron. J.
%%% \ao - Appl. Optics
% \apj - Astrophys. J.
% \apjl - Astrophys. J.
% \apjs - Astrophys. J. Suppl. Ser.
% \apss - Astrophys. Space Sci.
% \araa - Ann. Rev. Astron. Astrophys.
%%% \asr - Adv. Space Res.
% \astl - Astron. Lett.
% \atel - Astron. Telegram
% \azh - Астрон. журн.
% \baas - BAAS       
% \iauc - IAU Circ.
%%% \jrasc - JRASC     
% \mnras - Mon. Not. Roy. Astron. Soc.
%%% \nat{%%% \hbox{Nature
%%% \pra - Phys. Rev. A
%%% \prb - Phys. Rev. B
%%% \prc - Phys. Rev. C
%%% \prd - Phys. Rev. D
%%% \prl - Phys. Rev. Lett.    
%%% \pasp - Publ. Astron. Soc. Pacific
%%% \pasj - Publ. Astron. Soc. Japan
% \pazh - Письма в Астрон. журн.
%%% \qjras - QJRAS
%%% \skytel - S%%% \&T
%%% \solphys - Solar~Phys.
%\sovast - Sov.~Astron.
% \sval - Sov.~Astron. Lett.
%%% \ssr - Space~Sci. Rev.
%%% \zap - ZAp

%%% Список авторов задается командой \authors.
%%% В квадратных скобках размещается краткий список для верхнего колонтитула на четных страницах:
%%% ФАМИЛИЯ (заглавными буквами) - если автор только один;
%%% ФАМИЛИЯ1, ФАМИЛИЯ2 - если авторов двое;
%%% ФАМИЛИЯ1 и др. - если число авторов больше двух.
%%% В фигурных скобках размещается полный список авторов,
%%% причем каждый следующий добавляется командой \nextauth, имеющей два обязательных аргумента:
%%% в первых фигурных скобках указываются инициалы и фамилия, а во вторых - номера, соответствующие аффилиациям.
%%% Дополнительно, для основного соавтора в квадратных скобках перед фигурными указывается электронный адрес.

\authors[AFONINA et al.]{
  \nextauth[afonina.md19@physics.msu.ru]{M. D. Afonina}{1,2},
  \nextauth{A. V. Biryukov}{2,3,4},
  \nextauth[sergepolar@gmail.com]{S. B. Popov}{2, 5}
}

%%% Полный и краткий заголовки статьи задаются командой \titles.
%%% В квадратных скобках размещается краткий заголовок для верхнего колонтитула на нечетных страницах
%%% (для корректной печати этого колонтитула команда \titles должна располагаться после команды \authors).
%%% В фигурных скобках дается полное название.

\titles[Evolutionary status of long-period radio pulsars]{Evolutionary status of long-period radio pulsars}

%%% Список аффилиаций задается командой \affiliations, внутри которой каждая следующая аффилиация задается командой \nextaffil. 
%%% При этом последовательная нумерация аффилиаций производится автоматически.
%%% Однако, авторы должны сами проконтролировать соответствие этому списку номеров, указанных в командах \nextauth выше.

\affiliations{
\nextaffil{Department of Physics, Lomonosov Moscow State University, Moscow, Russia}
\nextaffil{Sternberg Astronomical Institute, Moscow, Russia }
\nextaffil{Faculty of Physics, HSE University, Moscow, Russia}
\nextaffil{Institute of Physics, Kazan Federal University, Kazan, Russia}
\nextaffil{ICTP~--~Abdus Salam International Center for Theoretical Physics, Trieste, Italy}
}

%%% Аннотация задается как аргумент команды \wideabstract. 
%%% В конце аннотации, внутри этой команды, задаются ключевые слова как аргумент команды \keywords.
%%% После них можно вставить "\doi{}", чтобы зарезервировать место для DOI.

\wideabstract{
We analyze the evolutionary status of recently discovered long-period radio sources
 PSR J0901-4046, GLEAM-X J1627-52, and GPM J1839-10. We discuss the hypothesis that all three sources are radio pulsars. In the framework of standard scenarios, it is often accepted that the pulsar mechanism is switched off when an external matter can penetrate the light cylinder. If the matter is stopped outside the light cylinder then the neutron star is at the ejector stage. We demonstrate that for realistic parameters of the interstellar medium, the 76-second pulsar PSR J0901-4046 might be at this stage. However, sources GLEAM-X J1627-52 and GPM J1839-10 with periods $\gtrsim 1000$~s can be ejectors only in the case of unrealistically large dipolar fields $\gtrsim 10^{16}$~G. Also, we show that neutron stars with spin periods $\sim 100$~s and dipolar magnetic fields $\lesssim 10^{13}$~G cannot be ejectors in a typical interstellar medium. Thus, we predict that long-period pulsars with standard fields will not be discovered.

%Мы рассматриваем эволюционный статус недавно обнаруженных долгопериодических радиоисточников  PSR J0901-4046, GLEAM-X J1627-52, GPM J1839-10. Существует предположение, что все три являются радиопульсарами. В рамках стандартных сценариев считается, что для работы пульсарного механизма необходимо исключить проникновение внешнего вещества под световой цилиндр, что соответствует стадии Эжектора. Мы показываем, что при реалистичных свойствах межзвездной среды 76-секундный пульсар PSR J0901-4046 должен находиться на этой стадии. В то время как источники GLEAM-X J1627-52 и GPM J1839-10 с периодами $\gtrsim 1000$~с могут находиться на этой стадии только при нереалистично высоких дипольных полях $\gtrsim 10^{16}$~Гс. Также мы показываем, что источники с периодами $\sim 100$~с и полями $\lesssim 10^{13}$~Гс не могут быть Эжекторами в реалистичной межзвездной среде.  Таким образом, предсказывается, что долгопериодические радиопульсары со стандартными магнитными полями не будут обнаружены. 

\keywords{neutron stars, radio pulsars}

%\doi{10.31857/S... (номер можно вставить, когда он станет известен)}
} 

%=========================================================

\section{Introduction}
\label{introduction}

Evolutionary status and observational appearances of isolated neutron stars (NSs) depend not only on intrinsic parameters of the compact objects (spin period, magnetic field, temperature, etc.) but also on interaction with the surrounding medium. This is mostly defined by two key parameters: density of the medium and NS relative velocity. 
Four main evolutionary stages of NSs are usually distinguished (see e.g., Lipunov 1992): ejector, propeller, accretor, and georotator. 
In this study, we discuss long-period radio pulsars, so the first two stages are of interest to us. 
At the ejector stage, the relativistic wind from the pulsar is strong enough to keep the external medium out of the light cylinder. 
The light cylinder radius corresponds to the maximal distance at which the magnetospheric field lines can exist co-rotating with the NS: 

\begin{equation}
    R_l = c/\omega.
    \label{rl}
\end{equation}
Here $c$~is the velocity of light, $\omega=2\pi / P$ is the spin frequency, and $P$~is the spin period.
At this stage, an NS can be observed as a radio pulsar. For the existence of such a source, it is necessary to have a cascade of electron-positron pair creation in the magnetosphere. Often, conditions for such cascade are defined in terms of a ``death line'' in the $P$~--~$\dot P$~diagram (see e.g., Beskin 1999). However, below we do not discuss these conditions as we are interested in a more general limitation related to the evolutionary status of NSs: we require that the NS is at the stage of an ejector.

%Эволюционный статус и наблюдательные проявления одиночных нейтронных звезд зависят не только от собственных параметров компактного объекта (период вращения, магнитное поле, температура и т.д.), но и от того, как происходит взаимодействие с внешней средой. Это определяется двумя основными параметрами: плотностью среды и скоростью нейтронной звезды относительно внешней среды.
% Выделяют четыре основных эволюционные стадии нейтронной звезды (см., например, Липунов 1987):
%Эжектор, Пропеллер, Аккретор и Георотатор. В данной работе, посвященной долгопериодическим радиопульсарам, нас будут интересовать первые две стадии. 
%На стадии Эжектора поток релятивистских частиц от нейтронной звезды является достаточно сильным, чтобы внешняя среда не попадала внутрь т.н. светового цилиндра. 
%Радиус светового цилиндра соответствует максимальному расстоянию, на котором возможно твердотельное вращение магнитных силовых линий:

%Здесь $c$~--- скорость света, $\omega=2\pi / P$~--- частота вращения, $P$~--- период вращения нейтронной звезды.
%Именно на этой стадии нейтронная звезда может проявлять себя как радиопульсар. Для этого необходимо, чтобы в магнитосфере работал каскад рождения электрон-позитронных пар. Часто условия существования каскада определяют через т.н. “линию смерти” (см., например, Бескин 1999). Нас, однако, в дальнейшем не будет интересовать выполнение этого условия, поскольку мы рассматриваем принципиальное ограничение (необходимое условие), связанное с нахождением нейтронной звезды на стадии эжекции. 

At the propeller stage, matter starts to penetrate inside the light cylinder preventing propagation of the relativistic wind and, finally, switching off the mechanism of its generation. 
Typically, one can define the condition for the transition from the ejector to the propeller stage as an equality of two characteristic radii. 
One of them is the gravitational capture radius (aka Bondi radius):

\begin{equation}
    R_G=\frac{2GM}{v^2}.
    \label{rG}
\end{equation}
Here $M$ is the NS mass, $v$~its velocity relative to the interstellar medium. 
Here and below we assume that $v$ is larger than the sound velocity in the interstellar medium: $c_s \sim \sqrt{kT/m_p} \sim 10$~km/s for $T \sim 10^4$~K (Klessen, Glover 2016).

%На стадии Пропеллера вещество может начать проникать под световой цилиндр, препятствуя распространению потока релятивистского ветра, в конечном счете “выключая” механизм его генерации. 
%В наиболее часто встречающихся случаях условие перехода со стадии Эжектора на стадию Пропеллера можно записать через равенство двух критических радиусов. 
%Одним из них является радиус гравитационного захвата:

%Здесь $M$~--- масса нейтронной звезды, а $v$~--- ее скорость относительно межзвездной среды. Причем здесь и далее мы считаем, что $v$ всегда больше скорости звука в теплой межзвездной среде: $c_s \sim \sqrt{kT/m_p} \sim 10$ км/с для $T \sim 10^4$ К (Клессен и Гловер 2016).

The second characteristic radius is the so-called Shvartsman radius. 
It can be derived from equating the pressure of relativistic pulsar wind and external pressure:

%Вторым критическим радиусом является т.н. радиус Шварцмана. Его можно выразить через равенство давления релятивистского (пульсарного) ветра и давления во внешней среде:

\begin{equation}
    R_{Sh}=\left(  \frac{\xi \mu^2 (GM)^2 \omega^4}{\dot M v^5 c^4} \right)^{1/2}.
    \label{rsh}
\end{equation}
Here $\mu=B\, R^3$~is the magnetic moment which can be defined by the equatorial magnetic field $B$ and neutron star radius $R$.
Parameter $\dot M$ expresses properties of the external medium. It is equal to the accretion rate when accretion is possible. We define this parameter as $\dot M= \pi R_G^2 \rho v$, where $\rho$~ is the density of the interstellar medium. Sometimes it is more convenient to use number density $n=\rho / m_p$, where $m_p$~is the proton mass. External pressure can be written as $\rho v^2$.

%Здесь $\mu=B\, R^3$~--- магнитный момент, который можно определить через поле на экваторе $B$ и радиус нейтронной звезды $R$. 
%Параметр $\dot M$ определяет свойства внешней среды и равен темпу аккреции, если таковая возможна. Мы оценивает его как $\dot M= \pi R_G^2 \rho v$, где $\rho$~--- плотность межзвездной среды. Иногда ее удобно выразить через концентрацию $n=\rho / m_p$, где $m_p$~--- масса протона. Давление внешней среды, соответственно, равно $\rho v^2$.

Eq.~(\ref{rsh}) is obtained using the following expression for the pulsar wind power:
$L_w = \xi \mu^2 \omega^4/c^4$, where the parameter $\xi \approx 1 + 1.4\sin^2\alpha$ depends on the angle  $\alpha$ between the spin axis and the magnetic dipole axis (Philippov et al. 2014). Under the assumption of isotropic and independent orientation of axes we obtain $\xi \approx 1.93$. Below for simplicity, we assume $\xi = 2$.

%Выражение (\ref{rsh}) основывается на том, что мощность пульсарного
%ветра равна $L_w = \xi \mu^2 \omega^4/c^4$, где
%фактор $\xi \approx 1 + 1.4\sin^2\alpha$ зависит от угла $\alpha$ между осью
%вращения и магнитной осью нейтронной звезды (Филиппов и др. 2014).
%В предположении изотропного распределени магнитных углов пульсаров
%$\xi \approx 1.93$. Далее в расчетах мы принимаем $\xi = 2$.

In the case when $R_G > R_l$ the ejector condition is formulated as $R_{Sh} > R_G$. 
However, for long spin periods, it can be that $R_l > R_G$. Then, the critical condition is written as $R_{Sh} > R_l$. 

%Условие существования Эжектора в том случае, когда $R_G > R_l$ формулируется как $R_{Sh} = R_G$. Однако для больших периодов вращения может сложиться ситуация, когда $R_l > R_G$. В таком случае критическим условием будет равенство $R_{Sh} = R_l$. 

For a given spin period, magnetic field, density of the surrounding medium, and NS velocity relative to this medium we can calculate if an NS is at the ejector stage or it is a propeller. In the first case, the compact object potentially can be a radio pulsar. But not in the second.

In the following section, we briefly describe the parameters of three recently discovered long-period radio sources and two similar objects known before. Then, in Sec.~3, we apply our considerations to the three long-period sources. In Sec.~4 we discuss our results and related subjects. Finally, in Sec.~5 we present our conclusions.

%Задав период вращения нейтронной звезды, ее магнитное поле, плотность внешней среды и скорость относительно нее, мы можем легко рассчитать, находится ли звезда на стадии Эжектора или Пропеллера. В первом случае возможно существование радиопульсара. Во втором~--- нет.
%В следующем разделе мы описываем параметры долгопериодических пульсаров и родственных им источников, а затем прилагаем эти простые соображения к недавно открытым долгопериодическим пульсарам.

%------------------------------------------------------
\section{Long-period pulsars}

For many years the longest periods of the known radio pulsars were of the order of 10~seconds. 
However, during the last few years, three radio sources with much longer periodicity have been discovered. In addition, two other objects possibly related to the long-period radio pulsars are known. In this section, we briefly describe the main observed properties of these sources.

%В течение долгого времени максимальные периоды известных радиопульсаров составляли $\sim 10$ секунд. Однако за последние два года было представлено три источника с существенно более длинными периодами. Кроме того, существует еще два интересных объекта, возможно родственных долгопериодическим пульсарам. В этом разделе мы кратко перечислим основные наблюдательные свойства этих источников.

PSR J0901-4046 was discovered in 2020 at the frequency 1.3 GHz with the MeerKAT radio telescope in South Africa (Caleb et al. 2022). It has the spin period 75.88~s and 
$\dot P=2.25\times 10^{-13}$~s/s. According to the standard magneto-dipole energy losses this corresponds to the surface dipolar field  $1.3 \times 10^{14}$~G. 
Shapes of different pulses are much more different from each other than it is typical for radio pulsars.
 %Due to a small dispersion measure ($52 \pm 1$ pc/cm$^{3}$), the distance to this pulsar is not well-measured. It is estimated as 330-470 pc depending on the applied model of the electron density distribution in the Galaxy.
 The small dispersion measure ($52 \pm 1$ pc/cm$^{3}$) corresponds to a distance $\sim$330-470 pc depending on the applied model of the electron density distribution in the Galaxy.

%%PSR J0901-4046; Caleb et al. 2022
% PSR J0901-4046 был открыт в 2020 году на радиотелескопе MeerKAT в Южной Африке на частоте 1.3 ГГц (Калеб и др. 2022). Он имеет период 75.88 секунды и $\dot P=2.25\times 10^{-13}$~с/c. По стандартной формуле для энергопотерь пульсаров это соответствует магнитному полю $1.3 \times 10^{14}$~Гс. 
%Форма его отдельных импульсов различна в гораздо большей степени, чем это обычно наблюдается у радиопульсаров. Вследствие малой меры дисперсии
%этого источика $52 \pm 1$ пк/см$^{3}$, расстояние до него 
%оценивается в 330-470 пк, в зависимости от модели плотности свободных
%электронов в Галактике.

The source GLEAM-X J1627-52 was discovered at low frequencies 72-231 MHz with the Murchison Widefield Array (MWA) in 2018 (Hurley-Walker et al. 2022).
 The phase of activity lasted for nearly three months. During this time just 71 pulses were detected. This gave an opportunity to identify the period 1091~s. 
For the period derivative, there is just an upper limit: $\dot P < (1-4) \times 10^{-9}$~s/s. 
Emission has high linear polarization ($\sim 88$\%). 
The brightness temperature is estimated as $10^{16}$~K. The radio luminosity exceeds the rotation energy losses by nearly three orders of magnitude. 

%Источник GLEAM-X J1627-52 был открыт  с помощью наблюдений на  радиотелескопе Murchison Widefield Array (MWA) в 2018 г. на частотах 72-231 МГц (Харлей-Уокер  и др. 2022). Активная фаза длилась около трех месяцев. 
%За это время был зарегистрирован 71 импульс, что позволило определить период 1091 с. 
%Для производной периода был получен предел $\dot P < (1-4) \times 10^{-9}$~с/c. Излучение обладает сильной линейной поляризацией ($\sim 88$\%). Яркостная температура оценивается в $10^{16}$~К. Радиосветимость примерно на три порядка превосходит потери вращательной энергии. 

GPM J1839-10 has been also discovered with the MWA (Hurley-Walker et al. 2023). 
Then, the source was observed with the Australia Telescope Compact Array (ATCA), the 
Parkes/Murriyang radio telescope, with the Australian Square Kilometre Array Pathfinder (ASKAP), and MeerKAT.
The period of pulsations is equal to 1318.2~s. The upper limit to the period derivative is $\dot P < 3 \times 10^{-9}$~s/s.
It is interesting that the source was also identified in old archival data that cover a period of over 30 years! The dispersion measure~--- $273.5 \pm 2.5$ pc cm$^{-3}$,~--- provides just a lower limit to the distance: $d \gtrsim 2.8$~kpc. 

%GPM J1839-10 также был впервые зарегистрирован на MWA (Харлей-Уокер и др. 2023). Затем источник наблюдался на Australia Telescope Compact Array (ATCA), радиотелескопе Parkes/Murriyang, на Australian Square Kilometre Array Pathfinder (ASKAP) и на MeerKAT. 
%Период пульсаций равен 1318.2 с. Предел на производную периода составляет $\dot P < 3 \times 10^{-9}$~с/c.
%Интересно, что источник удалось идентифицировать в архивных радиоданных, охватывающих более тридцати лет! Исходя из значения меры дисперсии $273.5 \pm 2.5$ пк см$^{-3}$, расстояние до GPM J1839-10
%удается, по сути, лишь ограничить снизу как $d \gtrsim 2.8$ кпк.

There is a source that in many respects reminds pulsating sources GLEAM-X J1627-52 and GPM J1839-10. This is the Galactic center radio transient GCRT J1745-3009. 
Initially, the source was detected by VLA at a low frequency of 0.33 GHz in 2002. Later it was detected a few times more (Hyman et al. 2005). During the first observed period of activity, the source demonstrated 5 bright ($\sim 1$~Jy) bursts with a typical duration of about 10 minutes. Intervals between the bursts were about 77 minutes. It was hypothesized that this could be a spin period of a compact object. 
During each of the consequent episodes of activity just a single bursts were detected, and they were significantly dimmer than the first five events. 
No counterparts were found in any spectral range. The source had a large brightness temperature which points towards a coherent emission mechanism. The nature of the object remains unclear (some exotic scenarios with NSs were discussed e.g., in Popov 2008). 

%Вероятно, есть источник, во многом напоминающий по своим свойствам пульсары GLEAM-X J1627-52 и GPM J1839-10. Это т.н. радиотранзиент в области центра Галактики (Galactic center radio transient~--- GCRT) GCRT J1745-3009. Источник  впервые был зарегистрирован в радиодиапазоне на VLA на частоте 0.33 ГГц в 2002 г., а затем наблюдался еще несколько раз (Хайман и др. 2005). Во время первой регистрации источник продемонстрировал пять мощных (порядка 1~Ян) всплесков длительностью около 10 минут. Интервал между всплесками составлял примерно 77 минут, что можно интерпретировать как вероятный период вращения. Во время последующих эпизодов активности было зарегистрировано лишь по одному всплеску, которые были слабее первых. Объект не удалось отождествить в других диапазонах спектра. Источник имел большую яркостную температуру, что указывает на когерентный механизм излучения. 
%Природа источника остается неизвестной (некоторые экзотические сценарии с нейтронными звездами обсуждались, например, в препринте Попов 2008).

Finally, let us mention the central source 1E161348-5055 in the supernova remnant RCW103. It was discovered in X-rays with the Einstein space observatory (Tuohy and Garmire 1980). It was a kind of sensation when very long-term periodic pulsations were discovered in this source (de Luca et al. 2008). The period of pulsations is 6.67 hours and the upper limit on the period derivative is $\dot P < 7 \times 10^{-10}$~s/s. Then, a magnetar activity of this source was discovered (Rea et al. 2016, D'Ai et al. 2016). For our discussion, this source is mainly interesting due to its long spin period and young age. This makes it a potential ``relative'' of GLEAM-X J1627-52 and GPM J1839-10 as they can have a similar mechanism of rapid initial spin-down (see Sec.~4 below) even though their present-day observational appearances are different.  
                                                     
%Наконец, упомянем источник 1E161348-5055 в остатке сверхновой RCW 103. Он был обнаружен обсерваторией Эйнштейн в рентгеновском диапазоне (Туохи, Гармире 1980). Важным результатом стало обнаружение периодических пульсаций излучения, связанных с вращением нейтронной звезды (Де Лука и др. 2008). Период вращения равен 6.67 часа, а производная периода имеет верхний предел $\dot P < 7 \times 10^{-10}$~с/c. Затем была обнаружена магнитарная активность этого объекта (Реа и др. 2016, Д'Аи и др. 2016). Нам этот источник интересен в первую очередь своим длинным периодом. Это делает его потенциальным ''родственником'' пульсаров GLEAM-X J1627-52 и GPM J1839-10, т.е. у них может быть общий механизм быстрого замедления (см. ниже раздел ''Обсуждение''), хотя современные наблюдательные проявления у них существенным образом различаются.

From the point of view of pulsar physics, long-period sources raise questions related to the emission mechanism. Firstly, all these sources are situated beyond the death line in the $P$--$\dot P$~diagram. Secondly, for some of them, even the radio luminosity is larger than the rotation energy losses. However, in the following section, we are going to discuss another problem related to the evolutionary status of these sources. As we will show, for some of the sources only an extreme (and so~--- unrealistic) combination of the values of magnetic field and spatial velocity brings them to the ejector stage.

%С точки зрения физики радиопульсаров объекты с большими периодами вызывают вопросы, связанные с механизмом генерации радиоизлучения. Во-первых, все они лежат за линией смерти.
%Во-вторых, у некоторых из них даже радиосветимость превосходит потери вращательной энергии. Однако в следующем разделе данной заметки мы рассмотрим совсем другую проблему, связанную с эволюционным статусом этих источников. Как мы увидим, для некоторых из них необходимо достаточно экстремальное сочетание величины магнитного поля и пространственной скорости для того, чтобы объекты находились на стадии эжекции.

\section{Results}

When $R_G>R_l$ from the equation $R_{Sh}=R_G$ one can obtain the following expression for the critical velocity:

%Из условия равенства $R_{Sh}=R_G$ при $R_G>R_l$ мы можем получить уравнение для критической скорости:
\begin{equation}
v_{p1} = \left(\dfrac{8\pi c^4 (GM)^2 \rho}{\mu^2\omega^4} \right)^{1/2}.
    \label{Rsh_RG}
\end{equation}

This expression can be rewritten as $v_{p1}=27.4 \, P_2^2 n^{1/2} B_{14}^{-1}$~km/s if the parameters of the NS are normalized to their typical values. Here, $P_2 = P/(100\mbox{~s})$, $B_{14}=B/(10^{14}\mbox{~G})$. %, and $n = n / (\mbox{~cm}^{-3})$.
 The mass of an NS is assumed to be $1.4\, M_\odot$, and its radius is $10$~km. If the velocity of an object is $v < v_{p1}$ then it is assumed to be at the propeller stage.

%Это уравнение можно переписать как $v_{p1}=27.4 \, P_2^2 n^{1/2} B_{14}^{-1}$~км/с, 
%где $P_2=P/100$~с, $B_{14}=B/10^{14}$~Гс.
%если параметры звезды нормированы на типичные (для нашей задачи)
%значения: период $P_2 = P/(100\mbox{~с})$, поле на экваторе $B_{14}=B/(10^{14}\mbox{~Гс})$, а концентрация $n$ берется в единицах см$^{-3}$. Масса звезды принята равной
%$1.4\, M_\odot$, радиус $10$ км. Если скорость объекта $v < v_{p1}$, то он оказывается на стадии Пропеллера.

In the case when $R_G < R_l$ it is necessary to use the condition $R_{Sh}=R_l$ to obtain the critical velocity:

%В случае же $R_G < R_l$ необходимо использовать условие $R_{Sh}=R_l$, и тогда для критической скорости получаем:

\begin{equation}
 v_{p2} = \left( \dfrac{\mu^2 \omega^6}{4\pi \rho c^6}\right)^{1/2}.
    \label{Rsh_Rl}
\end{equation}
After normalization the value of the critical velocity can be written as $v_{p2} = 2840 \, P_2^{-3} B_{14} n^{-1/2}$~km/s. The propeller stage corresponds to the velocity values $v > v_{p2}$.

%После аналогичных нормировок формула запишется как $v_{p2} = 2840 \, P_2^{-3} B_{14} n^{-1/2}$~км/с.
%Стадия Пропеллера в этом случае соответствует скорости выше критической: $v > v_{p2}$.

\begin{figure}[t]%[h!]
\centering
\includegraphics[width=\columnwidth]{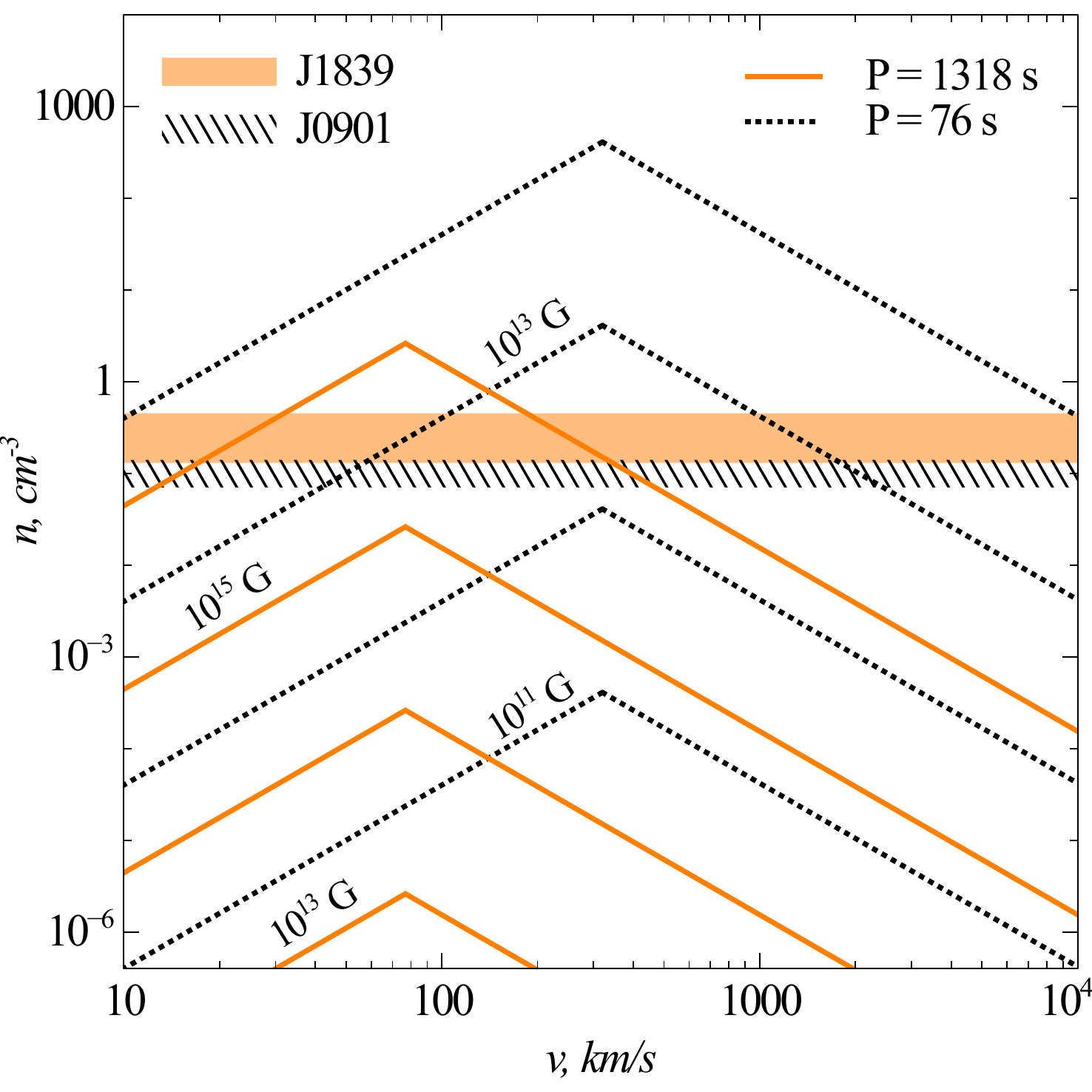}
\caption{The relationship between the critical velocity and the number density of the surrounding medium for the two objects PSR J0901-4046 ($P = 76$~s) and GPM J1839-10 ($P = 1318$~s). The region below each line corresponds to the ejector stage, the region above~--- to the propeller stage. Solid lines correspond to GPM J1839-10, while dashed lines~--- to PSR J0901-4046. For each period value, lines are drawn for several magnetic field values. The break in the lines corresponds to the velocity $v_{br}$. 
The left segments of each line correspond to eq.~(\ref{Rsh_RG}). Segments to the right from the break~--- to eq.~(\ref{Rsh_Rl}). The semi-transparent and hatched stripes show the local density estimates for GPM~J1839-10 and PSR~J0901-4046, respectively.}
\label{fig1}
\end{figure}

Using eqs.~(\ref{Rsh_RG}) and (\ref{Rsh_Rl}) we can define for which parameters the objects PSR J0901-4046, GLEAM-X J1627-52, and GPM J1839-10 can be at the ejector stage. 
In Fig.~\ref{fig1} we show a graphical representation of the equations for two values of the spin period~--- 76~s (black dashed lines) and 1318~s (orange solid lines)~--- and different values of the magnetic field.

%Рассмотрим, используя уравнения  (\ref{Rsh_RG}) и (\ref{Rsh_Rl}), при каких параметрах источники PSR J0901-4046, GLEAM-X J1627-52, GPM J1839-10 могут находиться на стадии эжекции. На Рис.~1 представлено графическое представление уравнений для разных значений магнитных полей и двух периодов: 76 секунд (черными прерывистыми линиями) и 1318 секунд (оранжевыми непрерывными линиями).

The break points in the plot correspond to the velocity $v_{br}$ defined by the condition $R_G=R_l$:

%Графики имеют излом на скорости $v_{br}$, определяемой из условия $R_G=R_l$:  
\begin{equation}
    v_{br} = \left( \dfrac{2GM\omega}{c}\right)^{1/2},
\end{equation}
or $v_{br} = 279 \, P_2^{-1/2}\mbox{~km/s}$ after normalisation.
%или $v_{br} = 279 \, P_2^{-1/2}\mbox{~км/с}$ после нормировки.

The region under each broken line corresponds to ejectors for given magnetic field values and number densities of the surrounding medium. Objects above the line are at the propeller stage. So, for example, for the period $P=1318$~s and magnetic field $B \lesssim 10^{14}$~G an NS can not be at the ejector stage if the number density $\gtrsim10^{-3.5}$~cm$^{-3}$.

%Область под каждой ломаной линией соответствует Эжекторам для данных значений магнитного поля и концентрации окружающей среды. Над линией находятся объекты на стадии Пропеллера. Так, например, для $P=1318$~с и полей $10^{14}$~Гс и ниже Эжекторов нет при концентрации среды $\gtrsim10^{-3.5}$~см$^{-3}$.

It is important to note that the magnetic field values shown in Fig.~\ref{fig1} are a kind of effective field. In the general case, these values $B$ can be related to the actual magnetic field values on the surface of the NS $B_0$ in the following way:

%Стоит, однако, отметить, что величины полей $B$, указанные рядом с соответствующими линиями на Рис.~1, до некоторой степени условны. В общем случае они могут быть связаны с реальным $B_0$ на поверхности звезды соотношением вида 
\begin{equation}
    \dfrac{B}{B_0} = R_{10}^{-3} \sin \alpha,
    \label{eq:Breal}
\end{equation}
where $R_{10} = R/(10\mbox{~km})$ is the NS radius. The first factor on the right-hand side of eq.~(\ref{eq:Breal}) reflects the fact that the real radii of NSs can be larger than the ``standard'' value of 10~km and is approximately equal to 11.5--12~km (e.g., Raaijmakers et al. 2021). The second multiplier reflects that the rotational energy losses of long-period radio pulsars may be closer to magneto-dipole losses ($\xi = \sin^2\alpha$) than to the classical pulsar losses ($\xi \approx 1 + 1.4\sin^2\alpha$). Since the cascade production of electron-positron pairs in the subpolar region of the NS is already terminated at periods $P_{d} \approx 16B_{14}^{8/15} \cos^{7/15}\alpha$~s (Novoselov et al. 2020), which is shorter than the period of any of the objects under discussion, each of them can formally be beyond its death line slowing down according to the magneto-dipole law. Accordingly, the condition for their transition from ejector to propeller may be milder (Beskin and Eliseeva 2005).

%где $R_{10} = R/(10\mbox{~км})$~--- радиус нейтронной звезды. Первый множитель в правой части (\ref{eq:Breal}) отражает то, что радиусы реальных нейтронных звёзд немного больше ''стандартного'' в 10 км и примерно равны 11.5--12 км (Рааймакерс и др. 2021). Второй же множитель связан с тем, что потери вращательной энергии долгопериодических пульсаров могут оказаться скорее близкими к магнитодипольным ($\xi = \sin^2\alpha$) нежели к классическим пульсарным ($\xi \approx 1 + 1.4\sin^2\alpha$). Так как каскадное рождение электрон-позитронных пар в приполярной области НЗ прекращается уже при периодах $P_{d} \approx 16B_{14}^{8/15} \cos^{7/15}\alpha$ с (Новосёлов и др. 2020), что меньше чем период любого из обсуждаемых объектов, то каждый из них может формально находится за своей линией смерти, замедляясь по магнитодипольному закону. Соответственно и условие их перехода с Эжектора на Пропеллер может быть более мягким (Бескин и Елисеева 2005).

Ultimately, each of the factors in eq.~(\ref{eq:Breal}) leads to an increase (by a factor of a few) in the value of the critical magnetic field at which the pulsar can no longer be an ejector at a given velocity and number density of the medium.
%В конечно итоге, каждый из множителей в (\ref{eq:Breal}) приводит к увеличению (в несколько раз) значения критического магнитного поля при котором пульсар уже не может при данной скорости и плотности среды быть Эжектором. 

%Соотношение между критической скоростью и концентрацией внешней среды  для двух объектов PSR J0901-4046 ($P = 76$~с) и GPM J1839-10 ($P = 1318$~с). Области под каждой из линий соответствуют стадии Эжектора, над~--- Пропеллера.  Непрерывные линии соответствуют GPM J1839-10. Пунктирные~--- PSR J0901-4046.  Для каждого значения периода линии нарисованы для нескольких значений напряженности магнитного поля. Излом в линиях соответствует скорости $v_{br}$. Слева от излома линия описывается уравнением (\ref{Rsh_RG}), справа~--- уравнением (\ref{Rsh_Rl}). Полупрозрачная и штрихованная полосы показывают оценки локальной плотности для GPM J1839-10 и PSR J0901-4046 соответственно.}
%\label{fig1}
%\end{figure}

In addition, the horizontal stripes (transparent and hatched) in Fig.~\ref{fig1} indicate estimates of the local density of the interstellar medium at the locations of each of the pulsars in the Galaxy. These estimates have been derived from a 3D map of the dust distribution in the Galaxy using combined data from Gaia Early Data Release 3 (Gaia EDR3) and the 2MASS catalogue (Vergely et al. 2022). This map allows estimating the differential absorption in the optical range $a_V$ (in magnitudes per parsec) in a volume of $10\times10\times0.8$~kpc centered on the Sun and with a spatial resolution of up to 10 pc. In this case, the total absorption (in magnitudes) in the direction of a given source is determined as the integral $A_V = \int a_V dl$ along the line of sight from the observer to the source.

%Кроме того, на графике горизонтальными полосами (полупрозрачной и 
%штрихованной) отмечены оценки для локальной плотности межзвездной 
%среды в тех местах Галактики, где находится каждый из пульсаров. Эти 
%оценки были получены на основе трехмерной карты распределения пыли в 
%Галактике по объединенным данным 3-го раннего релиза Gaia (Gaia EDR3) 
%и каталога 2MASS (Вергели и др. 2022). Эта карта позволяет оценить 
%дифференциальное поглощение в оптическом диапазоне
%$a_V$ (в зв. величинах на парсек) в объеме $10\times10\times0.8$ кпк
%c центром в Солнце и разрешением до 10 пк. 
%При этом полное поглощение (в звездных величинах) в направлении на заданный источник определяется как 
%интеграл по лучу зрения $A_V = \int a_V dl$
%от наблюдателя до источника.

In the direction of the PSR J0901-4046 in the distance interval of 330--470 pc from the Sun $a_V \sim 100-220\, \mu$mag/pc. From the X-ray observations it follows that the number density of atoms on the line of sight is proportional to the total absorption in the visible band $N_H = q\times A_V$, where $q\approx 2\times 10^{21}\,\mbox{cm}^{-2}~\mbox{mag}^{-1}$ (G{\"u}ver and {\"O}zel 2009). Therefore, for the given pulsar the local number density is $n = q \times a_V/(3.08\times 10^{18}\,\mbox{cm/pc}) \approx 0.07-0.14$~cm$^{-3}$. 

%В направлении на PSR J0901-4046 и интервале расстояний 330-470 пк от Солнца $a_V \sim 100-220\, \mu$mag/пк. Из наблюдений рентгеновских источников следует
%пропорциональность плотности атомов на луче зрения и полного
%поглощения в оптическом диапазоне $N_H = q\times A_V$, где $q\approx 2\times 10^{21}\,\mbox{cm}^{-2}~\mbox{mag}^{-1}$ (Гувер и Озел 2009).
%Поэтому для данного пульсара локальная плотность среды
%$n = q \times a_V/(3.08\times 10^{18}\,\mbox{см/пк}) \approx 0.07-0.14$~см$^{-3}$. 

For GPM J1839-10 the map from Vergely et al. (2022) provides a similar estimate only in the distance interval of 2.8--5~kpc, which is $n \approx 0.13-0.45$~cm$^{-3}$.

%Для GPM J1839-10 карта Вергели и др. (2022) позволяет сделать подобную оценку только в интервале расстояний 2.8-5 кпк, которая
%оказывается равной $n \approx 0.13-0.45$~см$^{-3}$.

The main conclusions from the analysis of Fig.~\ref{fig1} are as follows. First, PSR J0901-4046 with the magnetic field $\sim (1-2) \times 10^{14}$~G is at the ejector stage for almost all realistic values of the medium density and velocity. Second, the objects GLEAM-X J1627-52 and GPM J1839-10 can be at the ejector stage in a typical interstellar medium ($n\sim 0.1-1$ cm$^{-3}$) only with unrealistically high magnetic fields $\gtrsim 10^{16}$~G, or even higher if we take into account the corrections from eq.~(\ref{eq:Breal}). Finally, we can predict that the existence of pulsars with periods $\sim100$~s and magnetic fields $\lesssim 10^{13}$~G is practically impossible because it would require a very low density of the surrounding medium.

%Основные выводы, которые можно сделать из анализа Рис.~1 таковы. Во-первых, пульсар PSR J0901-4046 при магнитном поле $\sim (1-2) \times 10^{14}$~Гс практически при любых реалистичных значениях плотности среды и скорости находится на стадии Эжектора. Во-вторых, источники GLEAM-X J1627-52 и GPM J1839-10 могут находиться на стадии эжектора в типичной межзвездной среде ($n\sim 0.1-1$ см$^{-3}$) только при нереалистично больших магнитных полях $\gtrsim 10^{16}$~Гс или даже выше, если учесть поправки (\ref{eq:Breal}). Наконец, в-третьих, мы можем предсказать, что практически невозможно существование пульсаров с периодами $\sim100$ с и полями $\lesssim 10^{13}$~Гс, т.к. это потребовало бы очень низкой плотности окружающей их среды.

\section{Discussion}
% ф-ла 3 и далее содержит неопределенные коэффициенты порядка 1, что может сдвигать значения скоростей на десятки процентов

%Причина длинных периодов
% а) fallback
%Ronchi et al. 2022
%Beniamini et al. 2023
%б) эволюция в двойных

The origin of the long periods of the observed objects is currently unclear. It 
seems most likely that very long periods, such as those of GLEAM-X J1627-52 and 
GPM J1839-10, as well as of the source in the supernova remnant RCW 103, are 
associated with the fallback stage after a supernova explosion. This scenario 
is discussed in detail by Ronchi et al. (2022).  The population aspects of this 
scenario have been modeled by Rea et al. (2023) where the authors show that 
it is difficult to explain the origin of a large population of long-period 
radio pulsars within realistic assumptions made in this study.  

Another possibility, at least for the 76-second pulsar, seems to be the evolution in 
a massive close binary system, where the neutron star has time to reach the 
propeller or accretor stage before the second supernova explosion destroys the 
system. After the disruption of the binary, the older compact object is 
``reborn'' with a long spin period. This evolutionary path will be 
discussed in detail elsewhere (Kuranov and Popov, in prep.).   

If long-period pulsars experienced strong braking in the fallback accretion 
stage, they could enter the ejector stage from the propeller stage. In such a 
case, the critical condition would no longer be the equality $R_{Sh}=R_G$ or 
$R_{Sh}=R_l$. This is due to the so-called ``hysteresis effect'' (see 
Shvartsman 1970 and Lipunov 1992): the transition from the propeller to the ejector stage 
occurs with a shorter period (other parameters being equal) than the transition 
from ejector to propeller. Then, the transition condition is the equality of 
the magnetospheric radius $R_m$ and the radius of the light cylinder $R_l$.

The Alfvén radius can be used as the simplest estimate of the magnetospheric 
radius:

\begin{equation}
R_A=\left( \frac{\mu^2}{8\dot M \sqrt{2GM}} \right)^{2/7}.
    \label{RA}
\end{equation}
However, at the propeller stage, and even more so under the condition 
$R_m\approx R_l$, the magnetospheric radius can be much larger (see, e.g., 
Davis and Pringle 1981 and Lipunov 1987). At the propeller stage (when 
$R_m<R_G$) a good estimate is as follows:
\begin{equation}
%R_m= R_A^{7/9}\, R_G^{2/9}.
R_m = R_A \left(\dfrac{R_G}{R_A} \right)^{2/9}.
\label{rm1}
\end{equation}

If $R_m\approx R_l$ and $R_l>R_G$ then according to Davis and Pringle (1981)
one can write:

\begin{equation}
%R_m=2^{-1/6} \mu^{1/3}(GM)^{1/3} \dot M^{-1/6} v^{-5/6}.
R_m = \left( \dfrac{\mu^2 (GM)^2}{2\dot M v^5} \right)^{1/6}.
\label{rm2}
\end{equation}
This formula results from the equality of the magnetic pressure $\mu^2/(8 \pi 
R_m^{6})$ and the external pressure $\rho v^2$. 

However, it should be noted that at the time of the 
transition from the propeller to the ejector stage, the spin period should be 
slightly shorter than the present-day value, the magnetic field could be slightly 
higher (as it decays), and the parameters of the external environment may not 
correspond to the pulsar's current position. Nevertheless, for illustrative 
purposes, we present a plot similar to Fig.~1 using the equality $R_m=R_l$. The magnetospheric radius is calculated from eq.~(\ref{rm2}). In this 
case, the conditions do not depend on the relation between $R_G$ and $R_l$. So 
for the critical velocity we have:
\begin{equation}
v_{p3} = \left ( \dfrac{\mu^2\omega^6}{8\pi c^6 \rho} \right)^{1/2}.
%v_{crit3}= \frac{2^{3/2} \pi^{5/2} B\, R^3}{c^3 P^3 \rho^{1/2} } .
    \label{v3}
\end{equation}
In the normalized form, its value can be written as $v_{p3}= 1420 \, P_2^{-3} 
B_{14} n^{-1/2}$~km/s. If the velocity of an object exceeds $v_{p3}$, it is at 
the propeller stage. Note that, due to the ``hysteresis effect'' this is a 
tighter constraint on reaching the ejector stage than that given by eq.~(\ref{Rsh_Rl}).

\begin{figure}
\centering
\includegraphics[width=\columnwidth]{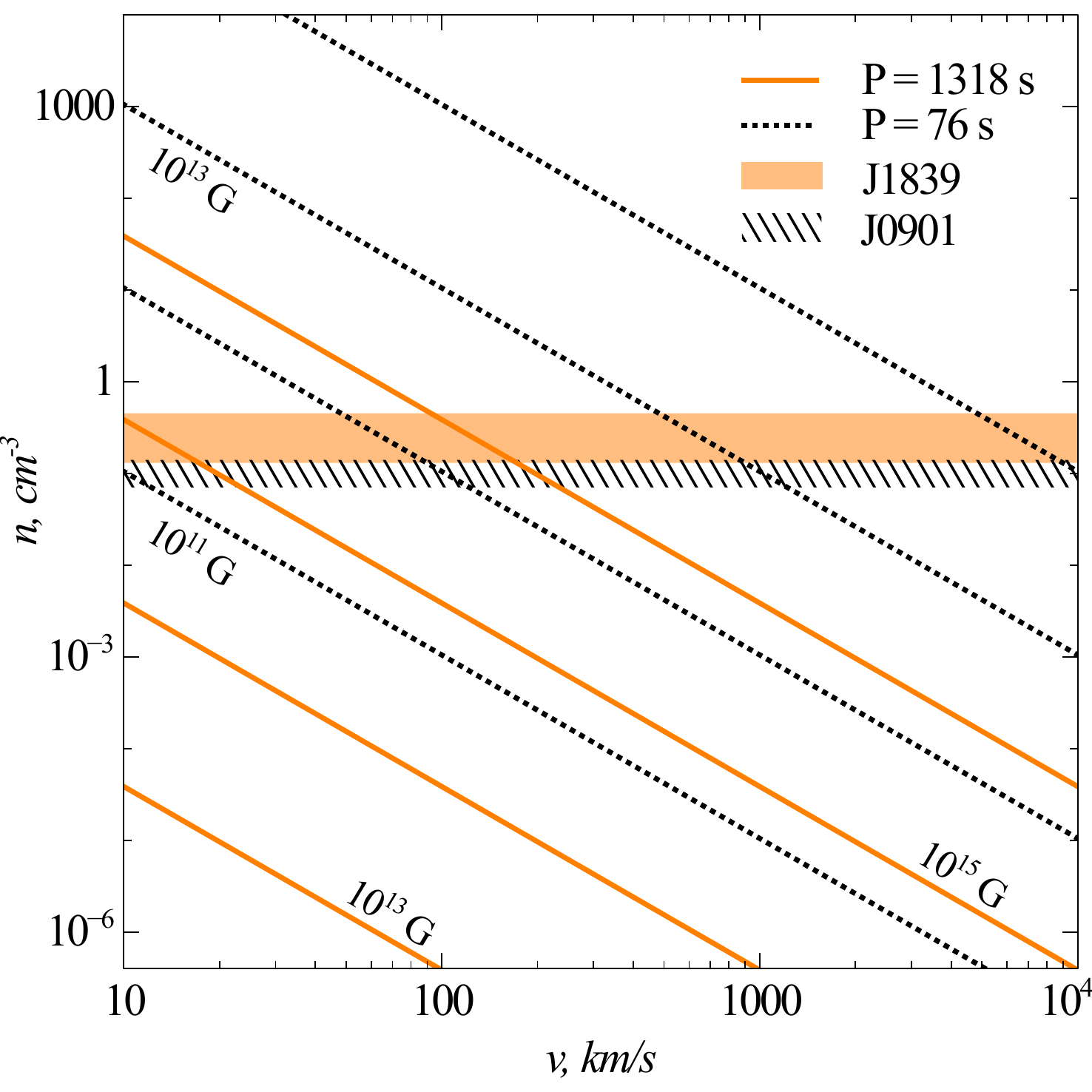}
\caption{The relationship between the critical velocity $v_{p3}$ and the 
density of the surrounding medium for two objects, PSR J0901-4046 ($P = 76$ s) 
and GPM J1839-10 ($P = 1318$ s), for four values of the magnetic field. The line styles and colours are the same as in Fig.~1.}
\label{fig2}
\end{figure}

Eq.~(\ref{v3}) is graphically represented in Fig.~2, similar to Fig.~1. Since it is not affected by the relationship between $R_G$ and 
$R_l$, we simply have a set of straight lines. The ejector region for a given 
field strength is below each corresponding line. 

Again, we see that the pulsar PSR J0901-4046 is in the ejector region. But the 
longer period sources (which apparently include GCRT J1745-3009) fall in the 
propeller region for realistic magnetic fields of $\lesssim10^{15}$~G at 
typical densities of the interstellar medium.  

%Magnetar-like activity can create a local cavern
Recall that the mechanism of radio emission in the sources GLEAM-X 
J1627-52 and GPM J1839-10 remains unknown. It is possible that this process is 
related not to the classical pulsar mechanism but to the magnetar mechanism. 
In particular, this is indirectly indicated by the fact that the radio 
luminosities of GLEAM-X J1627-52 and GPM J1839-10 exceed the rate of rotational 
energy loss. Therefore, another energy reservoir is required. This can 
reasonably be attributed to a strong magnetic field. In such a case, the 
activity of the NS can locally change the parameters of the external 
environment. Such a scenario should be considered in detail, but this is beyond 
the scope of this paper.  

%Выход на аккрецию
The future fate of long-period pulsars and related objects is an interesting 
question. By ``related objects'' we mean NSs with roughly the same 
parameters but different spatial velocities.  

If an NS has a long spin period (and possibly a strong magnetic field) 
early in its life, this will lead to its rapid transition to the accretion of interstellar matter. Moreover, at high velocities and with a large magnetic 
field, the object becomes a so-called georotator.\footnote{The detailed 
modeling of the evolution of isolated NSs with large initial spin periods, 
as well as the analysis of their properties at the accretion stage, will be presented by us in a separate 
publication (Afonina et al., in prep.).} Accordingly, the existence of a rather large population of isolated NSs that are able to start accreting interstellar matter in a time much shorter than the age 
of the Galaxy, should significantly increase estimates of the number of such sources. Of course, the population syntheses of isolated 
accreting NSs carried out so far (see e.g., Boldin and Popov 2010 and 
references therein) did not include such objects.

Finally, it is important to note that if the velocity and magnetic field 
distributions of long-period pulsars are similar to those of ordinary NSs then the vast majority of such objects will not be detected as 
normal radio pulsars. Thus, estimates of the number and birth rate of such objects based 
on radio observations alone may be significantly underestimated, since many 
young long-period NSs may be at the propeller stage.

%=========================================================
\section{Conclusions}
\label{sect:concl}

The discovery of long-period radio pulsars was an unexpected result. 
To date, there is no clear understanding of the nature of these objects, the 
emission mechanism, and the evolutionary path of these sources (see e.g., Rea et 
al. 2023).

We have considered constraints on the parameters of such sources when they are 
at the ejector stage within the framework of the pulsar model where the penetration 
of the external medium inside the light cylinder must be avoided to produce 
radio emission.

We show that the 76-second pulsar fully satisfies the requirements for being at the 
ejection stage. On the other hand, GLEAM-X J1627-52 and GPM J1839-10 
sources with spin periods of $\sim 10^3$ s cannot be ejectors in the standard 
interstellar medium unless their magnetic fields exceed  $\sim 10^{16}$~G, or they 
exhibit additional activity (e.g., magnetar) leading to a significant decrease 
of the matter density around them. Furthermore, if the rapid deceleration of rotation of 
these sources implies that they reached the propeller stage in the past, then the 
subsequent transition to the ejector stage may not be possible for realistic 
values of magnetic fields. We conclude that such long-period radio sources 
cannot be ordinary radio pulsars. 

In addition, we show that long-period pulsars with periods $\sim 
10^2$~s and fields $\lesssim 10^{13}$~G cannot be at the ejector stage in the standard interstellar medium. Thus, no analogs of PSR J0901-.4046 with a deceleration rate of $\dot P \lesssim 10^{-15}$~s/s will be detected.

\vskip 1cm

A. Biryukov thanks D. Wiebe for comments on the distribution of the interstellar medium in the Galaxy. M.A. and A.B. were supported by the RSF grant 21-12-00141.

\vskip 1cm

{\it Translated by the authors.}

\newpage

%Работа выполнена при поддержке.... (грант \textnumero\ .....).

%%% Список литературы предваряется командой \begin{references}, а завершается командой \end{references}.
%%% Библиографическое описание каждой ссылки должно предваряться командой \reference
%%% и должно соответствовать Правилам для авторов Писем в Астрономический журнал.
%%% Список литературы должен быть упорядочен по алфавиту.
%%% Для сокращенных наименований часто цитируемых журналов можно использовать команды:
%%% \aap - Astron. Astrophys.
%%% \aapr - Astron. Astrophys. Rev.
%%% \aaps - Astron. Astrophys. Suppl. Ser.
%%% \aj - Astron. J.
%%% \ao - Appl. Optics
%%% \apj - Astrophys. J.
%%% \apjl - Astrophys. J.
%%% \apjs - Astrophys. J. Suppl. Ser.
%%% \apss - Astrophys. Space Sci.
%%% \araa - Ann. Rev. Astron. Astrophys.
%%% \asr - Adv. Space Res.
%%% \astl - Astron. Lett.
%%% \atel - Astron. Telegram
%%% \azh - Астрон. журн.
%%% \baas - BAAS       
%%% \iauc - IAU Circ.
%%% \jrasc - JRASC     
%%% \memras - MmRAS
%%% \mnras - Mon. Not. Roy. Astron. Soc.
%%% \nat{%%% \hbox{Nature
%%% \pra - Phys. Rev. A
%%% \prb - Phys. Rev. B
%%% \prc - Phys. Rev. C
%%% \prd - Phys. Rev. D
%%% \prl - Phys. Rev. Lett.    
%%% \pasp - Publ. Astron. Soc. Pacific
%%% \pasj - Publ. Astron. Soc. Japan
%%% \pazh - Письма в Астрон. журн.
%%% \qjras - QJRAS
%%% \skytel - S%%% \&T
%%% \solphys - Solar~Phys.
%%% \sovast - Sov.~Astron.
%%% \sval - Sov.~Astron. Lett.
%%% \ssr - Space~Sci. Rev.
%%% \zap - ZAp

\end{document}